\pdfoutput=1

\documentclass[11pt]{article}

\usepackage[]{acl} 

\usepackage{times}
\usepackage{latexsym}

\usepackage{nicefrac}
\usepackage{xfrac}
\usepackage{tabularx, graphicx} 
\usepackage{booktabs}
\usepackage{subcaption}
\usepackage{soul}
\usepackage{amssymb}
\usepackage{pifont}
\usepackage{longtable,array}
\usepackage{lipsum} 
\usepackage{multirow}
\usepackage[bottom]{footmisc}

\usepackage[T1]{fontenc}

\usepackage[utf8]{inputenc}

\usepackage{microtype}

\usepackage{inconsolata}

\usepackage{graphicx}

\title{Online vs Offline: A Comparative Study of First-Party and Third-Party Evaluations of Social Chatbots}

\author{Ekaterina Svikhnushina \and Pearl Pu \\
        School of Computer and Communication Sciences \\ EPFL, Lausanne, Switzerland \\
        }

\begin{document}
\maketitle
\begin{abstract}
This paper explores the efficacy of online versus offline evaluation methods in assessing conversational chatbots, specifically comparing first-party direct interactions with third-party observational assessments. By extending a benchmarking dataset of user dialogs with empathetic chatbots with offline third-party evaluations, we present a systematic comparison between the feedback from online interactions and the more detached offline third-party evaluations. Our results reveal that offline human evaluations fail to capture the subtleties of human-chatbot interactions as effectively as online assessments. In comparison, automated third-party evaluations using a GPT-4 model offer a better approximation of first-party human judgments given detailed instructions. This study highlights the limitations of third-party evaluations in grasping the complexities of user experiences and advocates for the integration of direct interaction feedback in conversational AI evaluation to enhance system development and user satisfaction.
\end{abstract}

\section{Introduction}

Language modeling and conversational technologies continue to develop at an unprecedented pace. Evaluation of resulting models remains a challenge due to the lack of standardized approaches and the ever-growing scope of tasks and scenarios where Large Language Models (LLMs) are applied. Traditionally, both automatic and human evaluation metrics are applied to evaluate conversational agents \citep{10.1145/3641289}. Automatic evaluation is common and popular because it uses standard metrics and tools to evaluate model performance without intensive human involvement, saving time and minimizing subjective biases. On the other hand, human evaluation is considered more applicable to real-world scenarios where more comprehensive and nuanced feedback is necessary, especially in non-standard cases where automatic metrics may not suffice.

Approaches to human evaluation have been evolving over time, starting from designs where humans were asked to judge just one single generated model output to a static context \citep{vinyals2015neural} to designs where interactively generated dialogs became the focus of evaluation \citep{li2019acute}. Another important factor of consideration for human evaluation is whether it is conducted in an online or offline setting \citep{svikhnushina-pu-2023-approximating}. In the online setting, human workers initially interact with evaluated models and later provide their appraisals from the first-party perspective based on the immediate interaction experience. In the offline setting, conversational logs for evaluation are curated upfront and are provided to third-party judges for evaluation.

Several works have argued that online first-party evaluation of chatbots or language models serves as the most reasonable approximation of subjective human perceptions \citep{lee2022evaluating, svikhnushina-etal-2022-ieval, NEURIPS2019_fc981212}. However, to the best of our knowledge, no study has compared online and offline user annotation specifically for the task of conversational chatbot evaluation. We aim to address this gap in our work.

Overall, our contributions include the following. 1) We complement one of the existing datasets that features online human evaluations of chatbots with third-party scores for the same dialogs. 2) We juxtapose online and offline evaluation scores to draw conclusions about the two evaluation designs. 3) Finally, we compare offline human evaluation with an automatic evaluation, where the third-party scores are produced by a GPT-4 model. We discuss the implications of our results and intend to release the data from our experiments to facilitate future research endeavors.

\section{Related Work}

In the realm of human-computer interaction, the value of direct user feedback has been repeatedly highlighted. In their study of research-paper recommender systems, \citet{Beel2015} strongly support the necessity of involving users directly to gain accurate assessments of system effectiveness. Although their research centers on recommender systems, the insights provided might be relevant for other domains as well, emphasizing that user studies provide indispensable insights into how well a system aligns with user needs. This argument is further supported by \citet{jannach2023evaluating}, whose work specifically addresses conversational recommender systems, advocating strongly for the inclusion of user feedback and interaction in system evaluations.

The work by \citet{komatani-etal-2023-analyzing} closely aligns with our research, as both studies compare first- and third-party annotations in dialog interactions. Their analysis reveals significant perceptual differences between participants and external annotators, emphasizing the complexity involved in accurately capturing dialog experiences. While \citet{komatani-etal-2023-analyzing} focus on leveraging these insights to enhance machine learning model training, our study shifts the emphasis to evaluating the suitability of third-party annotations for assessing conversational systems.

\section{Materials and Methods}
To investigate whether third-party evaluation can serve as a reasonable approximation of first-party appraisals of a conversational chatbot, we propose running two identical evaluation experiments and comparing their results. In one experiment, first-party judgments should be collected from annotators in an interactive manner, where they first chat with the chatbots and provide their appraisal of the interaction experience afterwards. In the other experiment, the dialogs collected during the online interactive exchanges are used for offline evaluation: third-party annotators read static dialogs and use them as reference to evaluate quality of chatbot's responses.

\begin{table*}[]
\centering
\begin{tabular}{lll}
\toprule
                       & \multicolumn{1}{c}{Online}                              & \multicolumn{1}{c}{Offline}                            \\ \midrule
Main effect of chatbot     & $\text{F}_{3,1673}=257.92$, $p<0.001$* & $\text{F}_{3,1673}=85.12$, $p<0.001$* \\
Main effect of emotional polarity & $\text{F}_{1,1673}=43.17$, $p<0.001$*   & $\text{F}_{1,1673}=0.87$, $p=0.350$           \\
Interaction effect     & $\text{F}_{1,1673}=9.80$, $p<0.001$*  & $\text{F}_{1,1673}=1.43$, $p=0.232$          \\ \bottomrule
\end{tabular}
\caption{\label{tab:bemchmarking}
Comparison of ART ANOVA results. * indicates significant effect.
}
\end{table*}

\subsection{Dataset}
We decided to leverage the iEval dataset for our experiment \cite{svikhnushina-etal-2022-ieval}. To the best of our knowledge, this is the only publicly available dataset where the evaluation of chatbots was accomplished through interactive first-party annotations, i.e., the same human worker both interacted with the chatbots and then rated them. Other datasets featuring interactive exchanges between the human workers and the chatbots exist, e.g., DSTC9 \cite{gunasekara2020overview}, FED \cite{mehri-eskenazi-2020-unsupervised}. However, evaluations of chatbots in these datasets were arranged in an offline manner. Thus, they did not meet our requirements, whereas sticking to the iEval dataset, we could proceed directly to reusing the dialogs from the dataset for their evaluation in the offline setting.

The iEval dataset contains 1920 dialogues generated through interactions with four distinct dialogue models: Blender \cite{roller-etal-2021-recipes}, MIME \cite{majumder-etal-2020-mime}, MEED, and Plain \cite{xie-pu-2021-empathetic}. These dialogues were constructed around 480 unique emotional scenarios from the EmpatheticDialogues dataset \cite{rashkin-etal-2019-towards}, equally encompassing positive and negative emotional polarities. Participants were assigned an emotion and a situational description for each dialogue, then interacted with the chatbots over a structured six-turn conversation while role-playing the given scenario. Following the interactions, participants evaluated the chatbots on politeness, empathy, likability, repetitiveness, and making sense, using a 5-point Likert scale, in addition to providing an overall dialog quality rating on a 3-point scale. 

\subsection{Comparison of Results}
The first step in our comparison effort involves replicating the benchmarking methodology from the paper by \citet{svikhnushina-etal-2022-ieval} that introduced the iEval dataset. This consists of curating offline third-party annotations of the same dialogs under the same experimental design. The original research utilized a 2×4 within-subject factorial design to assess the impacts of chatbots and emotional polarity contexts (positive and negative), using the Aligned Rank Transform (ART) for non-parametric ANOVA \citep{Wobbrock2011}. Additionally, ordinal regression was used to explore how fine-grained factors influenced the rankings given to chatbots based on user perceptions. This approach will enable us to directly compare the outcomes of first-party and third-party annotations, determining whether these methodologies provide similar insights into chatbot performance.

As the second step, our study will employ the Dialog system Evaluation framework based on Prompting (DEP) \cite{svikhnushina-pu-2023-approximating}, an automated method designed to approximate online human evaluations of chatbots. We will analyze correlations between offline third-party human evaluations and online first-party quality annotations, and compare these to the correlations between DEP-based automatic evaluations and the same online annotations. This comparison will help determine which offline evaluation method -- human or automated -- more accurately mirrors online user feedback.

\section{Results}

\subsection{Annotation Experiment}
We ran the offline annotation task on Amazon Mechnical Turk, collecting annotations in batches between 24 October and 27 October 2023. The worker selection requirements followed the choices of the original work that introduced the iEval dataset. We aimed to replicate the experiment conditions as closely as possible to isolate the effect of interest (offline vs online setting). We required the workers to be US-based because the language of the annotation task was English. The workers spent an average of 7.6 minutes per assignment; the payment was set accordingly to exceed the US minimum wage standard. Following Mturk recommendations, we required the workers to have 98\% approval rate and 10,000 approved HITs. We further rejected the workers whose average assignment completion time or a number of contradictory responses to reverse-scaled questions in the Likert-type questionnaire raised concerns. For each annotation task from the original iEval dataset, 3 offline annotations were collected. Overall, 441 workers participated in our annotation effort providing their appraisals to the chatbots across 964 assignments, 244 of which were rejected (25.3\% rejection rate).

\begin{figure}[!b]
    \centering
    \includegraphics[width=\linewidth]{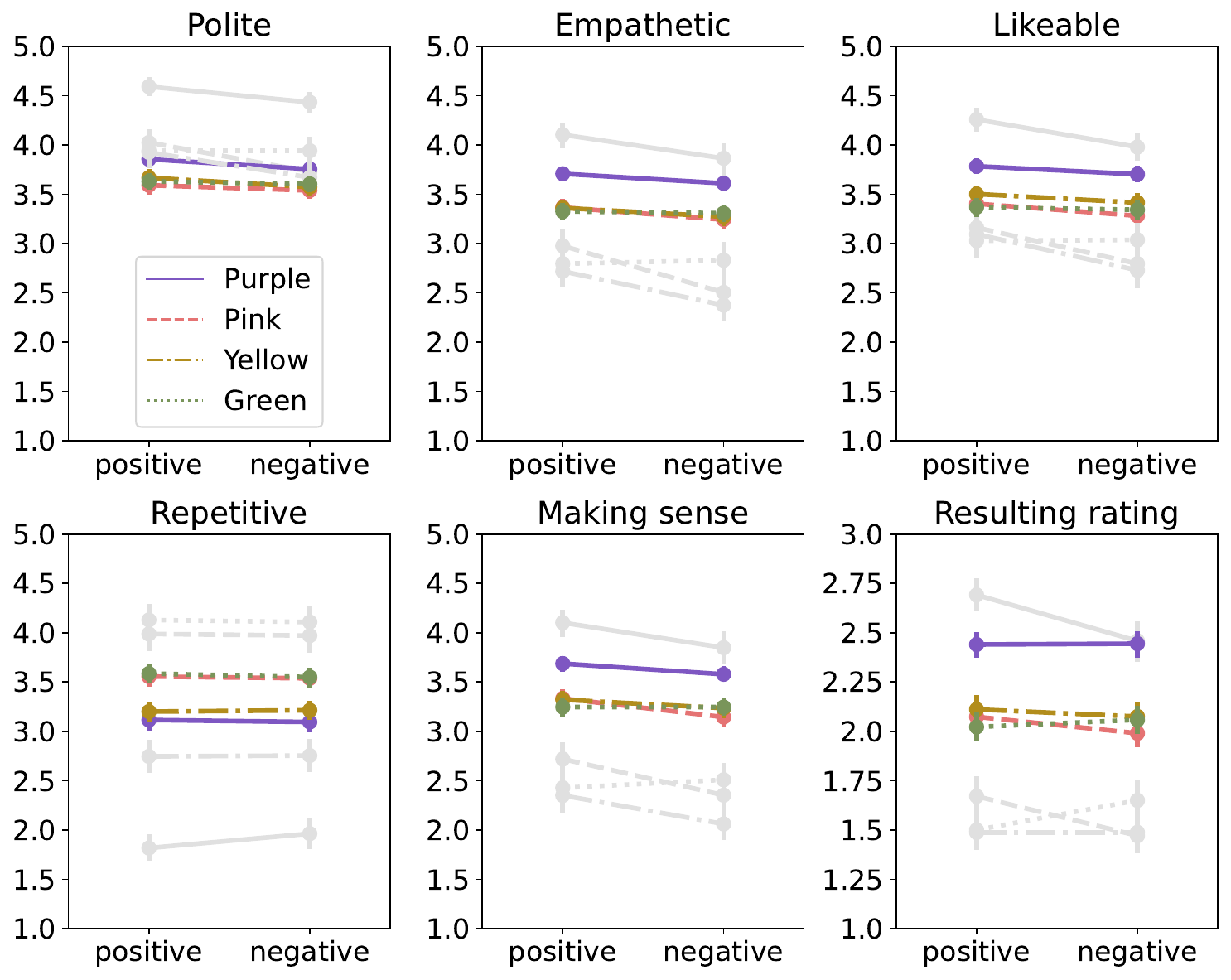}
    \caption{Benchmarking results of the four chatbots. Light-grey traces show the results from the online evaluation setup while colored lines represent the offline setup.}
    \label{fig:benchmarking}
\end{figure}

\subsection{Agreement Analysis for Offline Annotations}
We investigated agreement among third-party annotators for all evaluation dimensions using Krippendorff's alpha. Obtained values of the alpha statistic correspond to a slight level of agreement ($\alpha \leq 0.2$), ranging from 0.05 for the politeness dimension to 0.11 for the overall rating. For comparison, we also computed agreement scores for the evaluation dimensions of DSTC9 dataset \cite{gunasekara2020overview}: Krippendorff's alpha agreement values for all evaluation dimensions are around 0, indicating no agreement at all. Annotation scores in this dataset were collected in a similar offline fashion from third-party annotators. One of the reasons why the agreement scores are low might be the subjective nature of the evaluation task, leading to varying interpretations and perceptions among the readers of the dialogs.

\begin{figure}[!t]
    \centering
    \includegraphics[width=\linewidth]{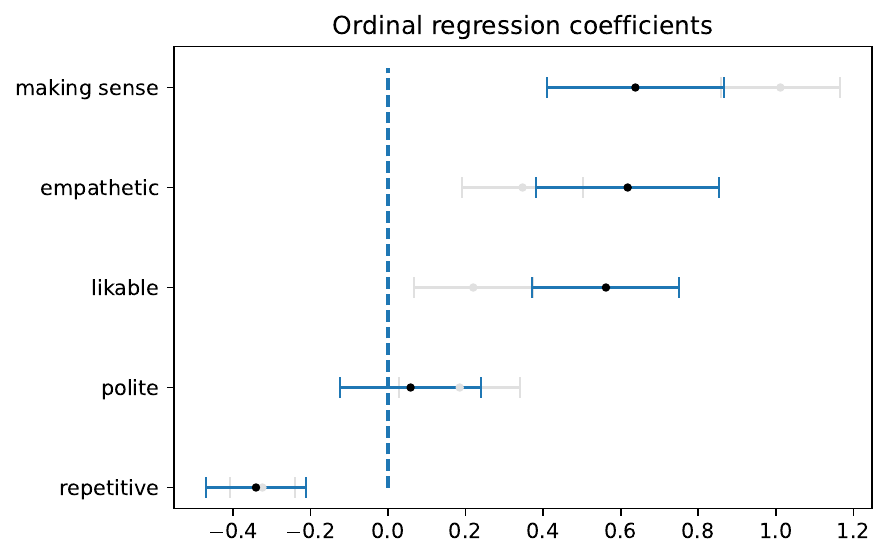}
    \caption{Results of ordinal regression on rank. 95\% confidence intervals are approximated as two standard errors. Light-grey traces show the results from the online evaluation setup while colored lines represent the offline setup.}
    \label{fig:regression}
\end{figure}

\subsection{Comparison of Benchmarking Results in Offline and Online Settings}

To compare the benchmarking of four chatbots, we used the non-parametric ART ANOVA procedure to analyze the rankings and juxtaposed the two results. The results are shown in Figure \ref{fig:benchmarking} and summarized in Table \ref{tab:bemchmarking}. Strikingly, the offline results lose the fine-grained information about evaluation outcomes. While online evaluation results indicated significant main effects of both chatbot, emotional polarity, and their interaction, offline results could only pinpoint significant difference of one chatbot.

Similar observations can be made considering the results of ordinal regression of various factors on the final ranking (Figure \ref{fig:regression}). Confidence intervals are considerably larger for the offline condition and some significant effects are lost.

\begin{table}[!b]
\centering
\resizebox{\linewidth}{!}{
\begin{tabular}{llll}
\toprule
                        &    & \multicolumn{1}{c}{Human} & \multicolumn{1}{c}{GPT-4} \\ \midrule
\multirow{2}{*}{Dialog:} & P: & 0.32, $p < 0.001$   & \textbf{0.59}, $p < 0.001$   \\
                        & S: & 0.32, $p<0.001$   & \textbf{0.58}, $p<0.001$   \\ \midrule
\multirow{2}{*}{System:} & P: & 0.97, $p<0.001$   & \textbf{0.99}, $p<0.001$   \\
                        & S: & 0.61, $p=0.11$              & \textbf{0.83}, $p<0.05$    \\ \bottomrule
\end{tabular}
}
\caption{\label{tab:correlation}
Correlation between first-party (online) human scores with two versions of third-party (offline) scores - human and automatic (prompted GPT-4 model). P and S stand for Pearson and Spearman correlation coefficients.
}
\end{table}

\subsection{Comparison of Human and Automatic Offline Evaluation}
\begin{figure}[!t]
    \centering
    \includegraphics[width=\linewidth]{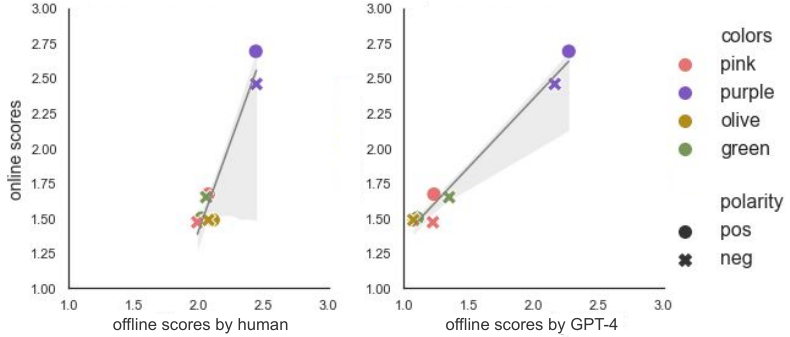}
    \caption{Scatter plots of system-level correlation.}
    \label{fig:correlation}
\end{figure}

Finally, we compared the result of offline evaluation results based on human annotations and automatic ones. To obtain automatic third-party scores, we prompted ChatGPT-4 model using the API\footnote{{\url{https://platform.openai.com/docs/models/gpt-4-turbo-and-gpt-4}}} providing it with the instructions from \cite{svikhnushina-pu-2023-approximating}. Results of third-party automatic evaluation consistently outperformed human third-party evaluation, both on dialog- and system-levels (Table \ref{tab:correlation}). System-level correlation plots are demonstrated in Figure \ref{fig:correlation}.

\section{Discussion and Conclusion}

Our study highlights significant disparities between first-party and third-party evaluations of open-domain empathetic chatbots, emphasizing that third-party assessments do not effectively mirror first-party user experiences. The low correlation and agreement among third-party raters underscore the challenge in capturing subjective perceptions through non-interactive assessments, particularly in more subjective and creative domains where direct interaction provides irreplaceable insights into user satisfaction and system effectiveness.

Furthermore, while third-party evaluations are less costly and logistically simpler, they fail to deliver the depth of understanding necessary for improving conversational agents in subjective applications. Our results advocate for prioritizing first-party evaluations to harness richer, more accurate user feedback. When cost constraints are predominant, leveraging automated evaluation methods, such as those utilizing advanced language models, could provide a balanced approach. It is essential to follow best practices of user simulation with LLMs \cite{10.1145/3637436} and ensure that instructions provided to these models are well-calibrated with first-party user perspectives \cite{svikhnushina-pu-2023-approximating}.

In conclusion, while automatic evaluations provide adequate measures for more objective dimensions, the nuanced nature of open-ended subjective tasks necessitates the inclusion of first-party user perspectives. These perspectives are critical in understanding and enhancing the user experience, where emotional and conversational subtleties play significant roles. Future research should focus on integrating user-centered evaluations with automated methodologies to optimize both the precision and practicality of assessments in conversational AI, ensuring that systems are effective across various tasks and aligned with user expectations.

\section*{Acknowledgments}
We would like to thank Antoine Bosselut for inspiring discussions to conceptualize this study and the Swiss National Science Foundation for funding this research (Grant No. 200021\textunderscore184602).

\bibliography{custom}

\begin{thebibliography}{18}
\providecommand{\natexlab}[1]{#1}

\bibitem[{Beel and Langer(2015)}]{Beel2015}
Joeran Beel and Stefan Langer. 2015.
\newblock A comparison of offline evaluations, online evaluations, and user studies in the context of research-paper recommender systems.
\newblock In \emph{Research and Advanced Technology for Digital Libraries}, pages 153--168, Cham. Springer International Publishing.

\bibitem[{Chang et~al.(2024)Chang, Wang, Wang, Wu, Yang, Zhu, Chen, Yi, Wang, Wang, Ye, Zhang, Chang, Yu, Yang, and Xie}]{10.1145/3641289}
Yupeng Chang, Xu~Wang, Jindong Wang, Yuan Wu, Linyi Yang, Kaijie Zhu, Hao Chen, Xiaoyuan Yi, Cunxiang Wang, Yidong Wang, Wei Ye, Yue Zhang, Yi~Chang, Philip~S. Yu, Qiang Yang, and Xing Xie. 2024.
\newblock \href {https://doi.org/10.1145/3641289} {A survey on evaluation of large language models}.
\newblock \emph{ACM Trans. Intell. Syst. Technol.}, 15(3).

\bibitem[{Ghandeharioun et~al.(2019)Ghandeharioun, Shen, Jaques, Ferguson, Jones, Lapedriza, and Picard}]{NEURIPS2019_fc981212}
Asma Ghandeharioun, Judy~Hanwen Shen, Natasha Jaques, Craig Ferguson, Noah Jones, Agata Lapedriza, and Rosalind Picard. 2019.
\newblock \href {https://proceedings.neurips.cc/paper_files/paper/2019/file/fc9812127bf09c7bd29ad6723c683fb5-Paper.pdf} {Approximating interactive human evaluation with self-play for open-domain dialog systems}.
\newblock In \emph{Advances in Neural Information Processing Systems}, volume~32. Curran Associates, Inc.

\bibitem[{Gunasekara et~al.(2020)Gunasekara, Kim, D'Haro, Rastogi, Chen, Eric, Hedayatnia, Gopalakrishnan, Liu, Huang, Hakkani-Tür, Li, Zhu, Luo, Liden, Huang, Shayandeh, Liang, Peng, Zhang, Shukla, Huang, Gao, Mehri, Feng, Gordon, Alavi, Traum, Eskenazi, Beirami, Eunjoon, Cho, Crook, De, Geramifard, Kottur, Moon, Poddar, and Subba}]{gunasekara2020overview}
Chulaka Gunasekara, Seokhwan Kim, Luis~Fernando D'Haro, Abhinav Rastogi, Yun-Nung Chen, Mihail Eric, Behnam Hedayatnia, Karthik Gopalakrishnan, Yang Liu, Chao-Wei Huang, Dilek Hakkani-Tür, Jinchao Li, Qi~Zhu, Lingxiao Luo, Lars Liden, Kaili Huang, Shahin Shayandeh, Runze Liang, Baolin Peng, Zheng Zhang, Swadheen Shukla, Minlie Huang, Jianfeng Gao, Shikib Mehri, Yulan Feng, Carla Gordon, Seyed~Hossein Alavi, David Traum, Maxine Eskenazi, Ahmad Beirami, Eunjoon, Cho, Paul~A. Crook, Ankita De, Alborz Geramifard, Satwik Kottur, Seungwhan Moon, Shivani Poddar, and Rajen Subba. 2020.
\newblock \href {https://arxiv.org/abs/2011.06486} {Overview of the ninth dialog system technology challenge: Dstc9}.
\newblock \emph{Preprint}, arXiv:2011.06486.

\bibitem[{Jannach(2023)}]{jannach2023evaluating}
Dietmar Jannach. 2023.
\newblock Evaluating conversational recommender systems: A landscape of research.
\newblock \emph{Artificial Intelligence Review}, 56(3):2365--2400.

\bibitem[{Komatani et~al.(2023)Komatani, Takeda, and Okada}]{komatani-etal-2023-analyzing}
Kazunori Komatani, Ryu Takeda, and Shogo Okada. 2023.
\newblock \href {https://doi.org/10.18653/v1/2023.sigdial-1.9} {Analyzing differences in subjective annotations by participants and third-party annotators in multimodal dialogue corpus}.
\newblock In \emph{Proceedings of the 24th Annual Meeting of the Special Interest Group on Discourse and Dialogue}, pages 104--113, Prague, Czechia. Association for Computational Linguistics.

\bibitem[{Lee et~al.(2022)Lee, Srivastava, Hardy, Thickstun, Durmus, Paranjape, Gerard-Ursin, Li, Ladhak, Rong, Wang, Kwon, Park, Cao, Lee, Bommasani, Bernstein, and Liang}]{lee2022evaluating}
Mina Lee, Megha Srivastava, Amelia Hardy, John Thickstun, Esin Durmus, Ashwin Paranjape, Ines Gerard-Ursin, Xiang~Lisa Li, Faisal Ladhak, Frieda Rong, Rose~E. Wang, Minae Kwon, Joon~Sung Park, Hancheng Cao, Tony Lee, Rishi Bommasani, Michael Bernstein, and Percy Liang. 2022.
\newblock \href {https://arxiv.org/abs/2212.09746} {Evaluating human-language model interaction}.
\newblock \emph{Preprint}, arXiv:2212.09746.

\bibitem[{Li et~al.(2019)Li, Weston, and Roller}]{li2019acute}
Margaret Li, Jason Weston, and Stephen Roller. 2019.
\newblock Acute-eval: Improved dialogue evaluation with optimized questions and multi-turn comparisons.
\newblock \emph{arXiv preprint arXiv:1909.03087}.

\bibitem[{Majumder et~al.(2020)Majumder, Hong, Peng, Lu, Ghosal, Gelbukh, Mihalcea, and Poria}]{majumder-etal-2020-mime}
Navonil Majumder, Pengfei Hong, Shanshan Peng, Jiankun Lu, Deepanway Ghosal, Alexander Gelbukh, Rada Mihalcea, and Soujanya Poria. 2020.
\newblock \href {https://doi.org/10.18653/v1/2020.emnlp-main.721} {{MIME}: {MIM}icking emotions for empathetic response generation}.
\newblock In \emph{Proceedings of the 2020 Conference on Empirical Methods in Natural Language Processing (EMNLP)}, pages 8968--8979, Online. Association for Computational Linguistics.

\bibitem[{Mehri and Eskenazi(2020)}]{mehri-eskenazi-2020-unsupervised}
Shikib Mehri and Maxine Eskenazi. 2020.
\newblock \href {https://aclanthology.org/2020.sigdial-1.28} {Unsupervised evaluation of interactive dialog with {D}ialo{GPT}}.
\newblock In \emph{Proceedings of the 21th Annual Meeting of the Special Interest Group on Discourse and Dialogue}, pages 225--235, 1st virtual meeting. Association for Computational Linguistics.

\bibitem[{Rashkin et~al.(2019)Rashkin, Smith, Li, and Boureau}]{rashkin-etal-2019-towards}
Hannah Rashkin, Eric~Michael Smith, Margaret Li, and Y-Lan Boureau. 2019.
\newblock \href {https://doi.org/10.18653/v1/P19-1534} {Towards empathetic open-domain conversation models: A new benchmark and dataset}.
\newblock In \emph{Proceedings of the 57th Annual Meeting of the Association for Computational Linguistics}, pages 5370--5381, Florence, Italy. Association for Computational Linguistics.

\bibitem[{Roller et~al.(2021)Roller, Dinan, Goyal, Ju, Williamson, Liu, Xu, Ott, Smith, Boureau, and Weston}]{roller-etal-2021-recipes}
Stephen Roller, Emily Dinan, Naman Goyal, Da~Ju, Mary Williamson, Yinhan Liu, Jing Xu, Myle Ott, Eric~Michael Smith, Y-Lan Boureau, and Jason Weston. 2021.
\newblock \href {https://doi.org/10.18653/v1/2021.eacl-main.24} {Recipes for building an open-domain chatbot}.
\newblock In \emph{Proceedings of the 16th Conference of the European Chapter of the Association for Computational Linguistics: Main Volume}, pages 300--325, Online. Association for Computational Linguistics.

\bibitem[{Schmidt et~al.(2024)Schmidt, Elagroudy, Draxler, Kreuter, and Welsch}]{10.1145/3637436}
Albrecht Schmidt, Passant Elagroudy, Fiona Draxler, Frauke Kreuter, and Robin Welsch. 2024.
\newblock \href {https://doi.org/10.1145/3637436} {Simulating the human in hcd with chatgpt: Redesigning interaction design with ai}.
\newblock \emph{Interactions}, 31(1):24–31.

\bibitem[{Svikhnushina et~al.(2022)Svikhnushina, Filippova, and Pu}]{svikhnushina-etal-2022-ieval}
Ekaterina Svikhnushina, Anastasiia Filippova, and Pearl Pu. 2022.
\newblock \href {https://aclanthology.org/2022.sigdial-1.41} {i{E}val: Interactive evaluation framework for open-domain empathetic chatbots}.
\newblock In \emph{Proceedings of the 23rd Annual Meeting of the Special Interest Group on Discourse and Dialogue}, pages 419--431, Edinburgh, UK. Association for Computational Linguistics.

\bibitem[{Svikhnushina and Pu(2023)}]{svikhnushina-pu-2023-approximating}
Ekaterina Svikhnushina and Pearl Pu. 2023.
\newblock \href {https://doi.org/10.18653/v1/2023.sigdial-1.25} {Approximating online human evaluation of social chatbots with prompting}.
\newblock In \emph{Proceedings of the 24th Annual Meeting of the Special Interest Group on Discourse and Dialogue}, pages 268--281, Prague, Czechia. Association for Computational Linguistics.

\bibitem[{Vinyals and Le(2015)}]{vinyals2015neural}
Oriol Vinyals and Quoc Le. 2015.
\newblock A neural conversational model.
\newblock \emph{arXiv preprint arXiv:1506.05869}.

\bibitem[{Wobbrock et~al.(2011)Wobbrock, Findlater, Gergle, and Higgins}]{Wobbrock2011}
Jacob~O. Wobbrock, Leah Findlater, Darren Gergle, and James~J. Higgins. 2011.
\newblock \href {https://doi.org/10.1145/1978942.1978963} {The aligned rank transform for nonparametric factorial analyses using only anova procedures}.
\newblock In \emph{Proceedings of the ACM Conference on Human Factors in Computing Systems (CHI '11)}, pages 143--146, New York. ACM Press.

\bibitem[{Xie and Pu(2021)}]{xie-pu-2021-empathetic}
Yubo Xie and Pearl Pu. 2021.
\newblock \href {https://doi.org/10.18653/v1/2021.conll-1.10} {Empathetic dialog generation with fine-grained intents}.
\newblock In \emph{Proceedings of the 25th Conference on Computational Natural Language Learning}, pages 133--147, Online. Association for Computational Linguistics.

\end{thebibliography}

\end{document}